\documentclass[pra, twocolumn, floatfix, superscriptaddress, longbibliography, preprintnumbers]{revtex4-1}

\usepackage{amsmath, amssymb, bbm, bbold}
\usepackage{graphicx}
\usepackage{color}

\newcommand{\ket}[1]{{| #1 \rangle}}
\newcommand{\bra}[1]{{\langle #1 |}}

\newcommand{\ii}{{\it i}}
\newcommand{\dd}{{\rm d}}

\newcommand{\sinc}{{\rm sinc}}

\newcommand{\FB}{{\rm f}}
\newcommand{\DB}{{\rm d}}
\newcommand{\aB}{{\rm a}}
\newcommand{\bB}{{\rm b}}
\newcommand{\hp}{{t_{\rm a b}}}
\newcommand{\lc}{a}

\begin{document}

\title{Lifetime of flatband states}

\author{Clemens Gneiting}
\email{clemens.gneiting@riken.jp}
\affiliation{Theoretical Quantum Physics Laboratory, RIKEN Cluster for Pioneering Research, Wako-shi, Saitama 351-0198, Japan}
\author{Zhou Li}
\affiliation{Theoretical Quantum Physics Laboratory, RIKEN Cluster for Pioneering Research, Wako-shi, Saitama 351-0198, Japan}
\author{Franco Nori}
\affiliation{Theoretical Quantum Physics Laboratory, RIKEN Cluster for Pioneering Research, Wako-shi, Saitama 351-0198, Japan}
\affiliation{Department of Physics, University of Michigan, Ann Arbor, Michigan 48109-1040, USA}

\date{\today}

\begin{abstract}
Flatbands feature the distortion-free storage of compact localized states of tailorable shape. Their reliable storage sojourn is, however, limited by disorder potentials, which generically cause uncontrolled coupling into dispersive bands. We find that, while detuning flatband states from band intersections suppresses their direct decay into dispersive bands, disorder-induced state distortion causes a delayed, dephasing-mediated decay, lifting the static nature of flatband states and setting a finite lifetime for the reliable storage sojourn. We exemplify this generic, disorder-induced decay mechanism at the cross-stitch lattice. Our analysis, which applies platform-independently, relies on the time-resolved treatment of disorder-averaged quantum systems with quantum master equations.
\end{abstract}

\preprint{\textsf{published in Phys.~Rev.~B~{\bf 98}, 134203 (2018)}}

\maketitle

\section{Introduction}

Flatbands, which may emerge as a consequence of symmetries or finetuning in certain tight-binding Hamiltonians, are characterized by a completely dispersionless single-particle energy spectrum, i.e., the band's energy $E(p)$ is independent of the Bloch state momentum $p$. Predicted several decades ago \cite{Sutherland1986localization, Lieb1989two}, they have recently become experimentally accessible in artificial lattice systems, ranging from electronic \cite{Nori1990angular, Vidal1998aharonov, Abilio1999magnetic, Drost2017topological, Slot2017experimental, Li2017realization, Cao2018unconventional}, to atomic \cite{Taie2015coherent, Ozawa2017interaction, Taie2017spatial} and photonic \cite{Nakata2012observation, Mukherjee2015observation, Vicencio2015observation, Kajiwara2016observation, Guzman2016experimental, Xia2016demonstration, Zong2016integrated, Baboux2016bosonic, Klembt2017polariton, Biondi2018emergent, Ozawa2018topological, Leykam2018artificial, Sun2018excitation}.

Remarkably, flatbands feature the existence of ``compact localized states'', free of any dynamical evolution and with tailorable shape, the latter by judiciously superposing the entirely degenerate Bloch states. Notably, these localized flatband states are even supported by a perfectly periodic lattice, whereas in standard dispersive bands localization usually emerges as a consequence of defects or disorder. Due to this localizability and absence of dispersive distortion, flatband states offer themselves as a means to store states and preserve information \cite{Bodyfelt2014flatbands, Rojas2017quantum}.

\begin{figure}[htb]
	\begin{center}
		\includegraphics[width=0.99 \linewidth]{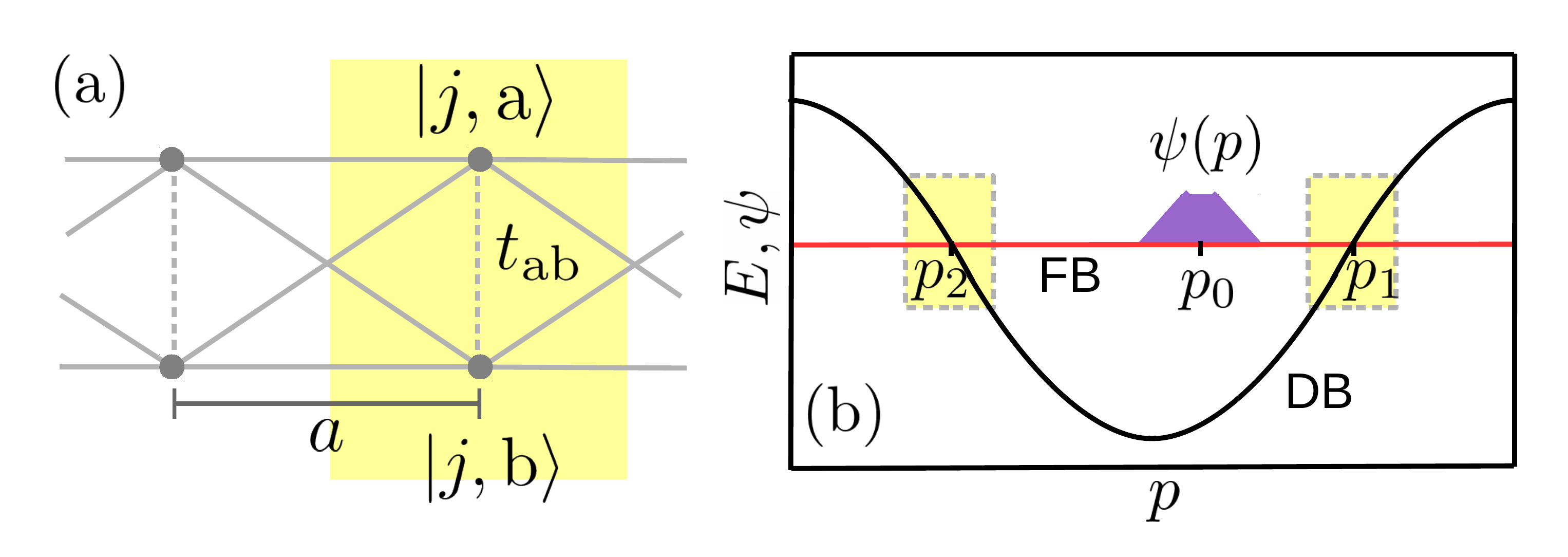}
	\end{center}
	\caption{\label{Fig:Cross-stitch_model} Cross-stitch lattice as paradigmatic flatband model. (a) It consists of two parallel sublattices, each unit cell $\ket{j}$ containing two sites $\ket{\aB}$ and $\ket{\bB}$ (yellow area). All neighboring sites are interconnected, i.e., hopping can occur within or by switching the sublattice. (b) The model exhibits two bands, one flat (FB) and one dispersive (DB). The intracell hopping $\hp$ controls their energetic relation. While an ideal flatband allows the distortion-free storage of compact localized states of tailorable shape, a disorder potential causes distortion and, in the vicinity of intersections (yellow areas), to a coupling into the dispersive band, limiting the state's reliable storage sojourn in the flatband.}
\end{figure}

The presence of disorder, however, may limit the static storage of flatband states. This is because a disorder potential, even if small, gives in general rise to spatially resolved phase fluctuations. While these may appear inconspicuous and initially irrelevant from the lattice perspective, they distort the wave packet in momentum space. In the vicinity of band intersections, this may eventually result in an uncontrolled coupling into dispersive bands (Fig.~\ref{Fig:Cross-stitch_model}), thus limiting the reliable storage sojourn, and ultimately resulting in the state's diffusive delocalization. In this sense, in flatband scenarios the reasoning is reversed: While localized in the perfectly periodic case, disorder, mediated by the coupling to a dispersive band, delocalizes flatband states.

In this paper, we study how disorder potentials induce the evolution of 1D flatband states in the presence of intersecting dispersive bands. This complements other studies on the impact of perturbations in flatband scenarios \cite{Goda2006inverse, Huber2010bose, Chalker2010anderson, Leykam2013flat, Bodyfelt2014flatbands, Leykam2017localization, Ge2017anomalous, Shukla2017criticality, Radoslavljevic2017light, An2017flux, Altfeder2017scanning}, e.g., describing the flatband-modified propagation in dispersive bands. We identify and characterize a generic, disorder-induced decay mechanism for flatband states, lifting their static nature and causing their effective diffusion, despite the absence of a kinetic term. We find that their (in)stability is controlled by the interplay of direct decay near intersections and dephasing-mediated state distortion. We demonstrate our findings with the cross-stitch lattice, which exhibits exactly one flat and one dispersive band (Fig.~\ref{Fig:Cross-stitch_model}) and therefore serves as a paradigmatic model system. Generic features, however, hold also in other (1D) flatband scenarios with band intersections, platform-independently. Our analysis relies on the treatment of disorder-averaged quantum systems with quantum master equations \cite{Gneiting2016incoherent, Kropf2016effective, Gneiting2017quantum, Gneiting2017disorder, Chen2017simulating, Gneiting2018disorder}.

\section{Single-intersection approximation}

To motivate our ansatz Hamiltonian (\ref{Eq:single_intersection_model}), we derive it now from the quasi-onedimensional cross-stitch lattice, which is composed of two parallel sublattices $\ket{\aB}$ and $\ket{\bB}$ with intra- and interlattice nearest neighbor hopping, see Fig.~\ref{Fig:Cross-stitch_model}. The Hamiltonian (in the absence of a potential) reads (e.g., \cite{Flach2014detangling}) $\hat{H} =$
\begin{align} \label{Eq:cross-stitch_Hamiltonian}
-J \sum_{j \in \mathbb{Z}} \Big\{& (\ket{j}\bra{j+1} + \ket{j}\bra{j-1}) \otimes (\mathbb{1}_2 + \ket{\aB}\bra{\bB} + \ket{\bB}\bra{\aB}) \nonumber \\
&+\hp \ket{j}\bra{j} \otimes (\ket{\aB}\bra{\bB} + \ket{\bB}\bra{\aB}) \Big\} ,
\end{align}
exhibiting two bands, one flat, $E_\FB = J \hp$, and one dispersive, $E_\DB(k) = -4 J \cos(k) - J \hp$, with hopping constant $J$ and intracell hopping participation $\hp$. The bands intersect twice if $|\hp| < 2$, which we assume from now on. States in a symmetric superposition of the two sublattices reside in the dispersive band $\ket{\DB}$, while antisymmetric superpositions reside in the flatband $\ket{\FB}$, $\ket{\FB} = (\ket{\aB} - \ket{\bB})/\sqrt{2}$ and $\ket{\DB} = (\ket{\aB} + \ket{\bB})/\sqrt{2}$. The Hamiltonian (\ref{Eq:cross-stitch_Hamiltonian}) then reads $\hat{H} = -J (4 \cos[\hat{p} \lc/\hbar]+ \hp) \ket{\DB}\bra{\DB} + \hp J \ket{\FB}\bra{\FB}$, with lattice constant $\lc$.

As the vicinities of the intersections dominate the decay of flatband states [Fig.~\ref{Fig:Cross-stitch_model}(b)], we hereafter linearize the dispersive band at the intersection $p_1$ closest to the flatband state. Below, we will reintroduce the second intersection. Without loss of generality, we assume $E_\FB = 0$. Moreover, we assume that the flatband state extends over at least a few unit cells, legitimating the continuum limit. The Hamiltonian (\ref{Eq:cross-stitch_Hamiltonian}) is then approximated by
\begin{align} \label{Eq:single_intersection_model}
\hat{H} = v (\hat{p}-p_1) \otimes \ket{\DB}\bra{\DB} ,
\end{align}
with $[\hat{x},\hat{p}]=i \hbar$, and the velocity $v$ the dispersive-band slope at the intersection.

In a perfect implementation of (\ref{Eq:cross-stitch_Hamiltonian}) (or (\ref{Eq:single_intersection_model}), respectively), a state residing in the flatband would not evolve. More realistically, however, one should consider at least small potential variations, e.g., due to impurities or stray fields. A general disorder potential in the cross-stitch model is written (again in the continuum limit)
\begin{align} \label{Eq:Disorder_potential}
\hat{V}_{\varepsilon} = V_{\varepsilon}^{\aB}(\hat{x}) \otimes \ket{\aB}\bra{\aB} + V_{\varepsilon}^{\bB}(\hat{x}) \otimes \ket{\bB}\bra{\bB} ,
\end{align}
where $\epsilon$ labels different disorder realizations and can be discrete, continuous and/or a multi-index (for convenience, we write integrals throughout). We assume that the disorder potential vanishes on average, $\int \dd \varepsilon \, p_{\varepsilon} \hat{V}_{\varepsilon} = 0$ ($p_{\varepsilon}$ denotes the probability distribution over the disorder realizations), i.e., the full Hamiltonian is $\hat{H}_{\varepsilon} = \hat{\overline{H}} + \hat{V}_{\varepsilon}$ with the average Hamiltonian $\hat{\overline{H}}$ as in (\ref{Eq:single_intersection_model}). Moreover, we assume that the disorder potential is weak, such that only dispersive band states in the vicinity of the intersection, where the linear band approximation is valid, become accessible.

The two sublattices $\ket{\aB}$ and $\ket{\bB}$ in general exhibit differing, but correlated disorder potentials $V_{\varepsilon}^{\aB}(x)$ and $V_{\varepsilon}^{\bB}(x)$. Assuming homogeneous disorder, we define the intra- and inter-sublattice correlations
\begin{align} \label{Eq:Correlation_functions}
C_{\sigma \sigma'}(x-x') &\equiv \int \dd \varepsilon \, p_{\varepsilon}  V_{\varepsilon}^{\sigma}(x) V_{\varepsilon}^{\sigma'}(x') \nonumber \\
 &= \int \dd q \, e^{\ii q (x-x')/\hbar} \, G_{\sigma \sigma'}(q) ,
\end{align}
with $\sigma, \sigma' \in \{ \aB, \bB \}$, and $G_{\sigma \sigma'}(q)$ describing the disorder-induced scattering. The intersublattice disorder correlations strongly influence the disorder-induced band coupling. Indeed, rewriting the disorder potential \cite{Flach2014detangling}, $\hat{V}_{\varepsilon} = V_{\varepsilon}^{+}(\hat{x}) \otimes \mathbb{1}_2 + V_{\varepsilon}^{-}(\hat{x}) \otimes \hat{\sigma}_x$, with $\hat{\sigma}_x \equiv \ket{\FB}\bra{\DB} + \ket{\DB}\bra{\FB}$ and $V_{\varepsilon}^{\pm}(x) = \frac{1}{2} \left[ V_{\varepsilon}^{\aB}(x) \pm V_{\varepsilon}^{\bB}(x) \right]$, reveals that the interband coupling, mediated by $V_{\varepsilon}^{-}(x)$, vanishes if $V_{\varepsilon}^{\aB}(x) = V_{\varepsilon}^{\bB}(x)$.

For simplicity, we assume that the intrasublattice correlations are the same on both sublattices, $G_{\sigma \sigma'}(q) = G_{\sigma-\sigma'}(q)$ (for convenience, we replace $\aB \rightarrow 1/2$ and $\bB \rightarrow -1/2$, i.e., $(\sigma-\sigma') \in \{-1,0,1\}$). We then define
\begin{align} \label{Eq:Internal_diagonalization}
G_{\sigma-\sigma'}(q) = \sum_{\beta \in \{ -1,0,1 \} } e^{\ii \pi \beta (\sigma-\sigma')}\tilde{G}_{\beta}(q) ,
\end{align}
with [$G_{-1}(q)=G_1(q)$] $\tilde{G}_{0}(q) = \frac{1}{2} [G_0(q)+G_1(q)]$ and $\tilde{G}_{1}(q) = \tilde{G}_{-1}(q) = \frac{1}{4} [G_0(q)-G_1(q)]$. Finally, as they originate from the same disorder potentials, we require that the intersublattice and the intrasublattice correlations have the same form, generically modified by a factor $-1 \leq \delta \leq 1$:
\begin{align} \label{Eq:Consistency_condition}
C_{\aB \bB}(x) = \delta \, C_{\aB \aB}(x) = \delta \, C_{\bB \bB}(x) .
\end{align}
This yields $\tilde{G}_{0}(q) = G_0(q) (1+\delta)/2$ and $\tilde{G}_{1}(q) = G_0(q) (1-\delta)/4$. Note that $\delta=+1$ ($\delta=-1$) describes perfectly correlated (anticorrelated) sublattice potentials, while intermediate $\delta$ values can, e.g., result from their weighted combination, corresponding to several disorder sources, some causing correlated, some anticorrelated, disorder potentials on the sublattices. We remark that, if the intrasublattice correlations differ, Eq.~(\ref{Eq:Consistency_condition}) does in general not hold. While such generalization is feasible within our framework, no additional insights would emerge.

Note that (\ref{Eq:single_intersection_model}) is {\it not} limited to the cross-stitch lattice, but serves as a generic, linearized model of {\it any} flatband-dispersive band intersection. In general, $\ket{\aB}$ and $\ket{\bB}$ then correspond to unspecified internal states of the unit cell, with $(\ket{\FB}, \ket{\DB})^{T} = U (\ket{\aB}, \ket{\bB})^{T}$. For our discussion, it suffices to focus on disorder potentials $\hat{V}_{\varepsilon} = V_{\varepsilon}^{\rm intra}(\hat{x}) \otimes \mathbb{1}_2 + V_{\varepsilon}^{\rm inter}(\hat{x}) \otimes \hat{\sigma}_x$, as in the cross-stitch model; $\hat{\sigma}_y$ and $\hat{\sigma}_z$ contributions could, however, easily be included. Disorder-induced modifications of the kinetic term (\ref{Eq:single_intersection_model}) are also neglected here. For clarity, we continue to discuss the cross-stitch lattice.

\section{Disorder-averaged time evolution}

We now describe the time evolution of the disorder-averaged quantum state $\overline{\rho}(t) = \int \dd \varepsilon \, p_{\varepsilon} \, \rho_{\varepsilon}(t)$, where $\rho_{\varepsilon}(t) = e^{-i \hat{H}_{\varepsilon} t/\hbar} \rho_0 e^{i \hat{H}_{\varepsilon} t/\hbar}$. This is achieved with a quantum master equation perturbative in the disorder potential \cite{Gneiting2017quantum, Gneiting2017disorder}. Using (\ref{Eq:Correlation_functions}) and (\ref{Eq:Internal_diagonalization}), one obtains the time-local, translation-covariant master equation $\partial_t \overline{\rho}(t)$
\begin{subequations} \label{Eq:Translation-covariant_master_equation}
\begin{align}
=& -\frac{\ii}{\hbar} [\hat{H}_{\rm eff}(t), \overline{\rho}(t)] \\
+& \!\! \sum_{\alpha \in \{ \pm 1 \}} \frac{2 \alpha}{\hbar^2} \int_{-\infty}^{\infty} \!\! \dd q \!\! \sum_{\beta \in \{ -1,0,1 \} } \!\!\!\!\!\! \tilde{G}_{\beta}(q) \int_{0}^{t} \!\! \dd t' \mathcal{L} \left( \hat{L}^{(\alpha)}_{q,\beta}(t'), \overline{\rho}(t) \right) , \nonumber
\end{align}
where $\mathcal{L}(\hat{L},\rho) = \hat{L} \rho \hat{L}^{\dagger} - \frac{1}{2} \hat{L}^{\dagger}\hat{L} \rho - \frac{1}{2} \rho \hat{L}^{\dagger}\hat{L}$ and $G_\beta(-q) = G_\beta(q)$. The effective Hamiltonian $\hat{H}_{\rm eff}(t) = \hat{H}_{\rm eff}^{\dagger}(t)$ and the Lindblad operators $\hat{L}^{(\alpha)}_{q,\beta}(t)$ are given by $\hat{H}_{\rm eff}(t)$
\begin{align}
= \hat{\overline{H}} - \frac{\ii}{2 \hbar} &\int_{-\infty}^{\infty} \!\! \dd q \!\! \sum_{\beta \in \{ -1,0,1 \} } \!\!\!\!\!\! \tilde{G}_{\beta}(q) \int_{0}^{t} \!\! \dd t' \, [\hat{V}_{q,\beta},\hat{\tilde{V}}_{-q,-\beta}(t')] , \nonumber \\
&\hat{L}^{(\alpha)}_{q,\beta}(t) = \frac{1}{2} \left[ \hat{V}_{q,\beta} + \alpha \hat{\tilde{V}}_{q,\beta}(t) \right] ,
\end{align}
\end{subequations}
where $\hat{\tilde{V}}_{q,\beta}(t) = e^{-\ii \hat{\overline{H}} t/\hbar} \, \hat{V}_{q,\beta} \, e^{\ii \hat{\overline{H}} t/\hbar}$ and $\hat{V}_{q,\beta} = e^{\ii q \hat{x}/\hbar} \otimes e^{\ii (\pi/2) \beta \hat{\sigma}_x}$. With $\hat{\overline{H}} = v (\hat{p}-p_1) \otimes \ket{\DB}\bra{\DB}$ [cf.~Eq.~(\ref{Eq:single_intersection_model})], we then have $\hat{\tilde{V}}_{q,\beta=0}(t) = e^{\ii q \hat{x}/\hbar} \otimes \left( \ket{\FB}\bra{\FB} + e^{-\ii v t q/\hbar} \ket{\DB}\bra{\DB} \right)$, $\hat{\tilde{V}}_{q,\beta=1}(t) = \ii e^{-\ii v t (\hat{p}-p_1)/\hbar} e^{\ii q \hat{x}/\hbar} \otimes \ket{\DB}\bra{\FB} + \ii e^{\ii q \hat{x}/\hbar} e^{\ii v t (\hat{p}-p_1)/\hbar} \otimes \ket{\FB}\bra{\DB}$, and $\hat{\tilde{V}}_{q,\beta=-1}(t) = -\hat{\tilde{V}}_{q,\beta=1}(t)$.

The master equation (\ref{Eq:Translation-covariant_master_equation}) describes the disorder-perturbed evolution of the full two-band quantum state. In the following, we focus on the flatband component $\overline{\rho}_{\FB} \equiv \bra{\FB}\overline{\rho}\ket{\FB}$. Projecting (\ref{Eq:Translation-covariant_master_equation}) onto $\ket{\FB}$, we obtain ($\overline{\rho}_{\DB} = \bra{\DB}\overline{\rho}\ket{\DB}$)
\begin{subequations} \label{Eq:Flatband_evolution}
\begin{align}
\partial_t \overline{\rho}_{\FB} &= \frac{2 t}{\hbar^2} \int_{-\infty}^{\infty} \!\! \dd q \, \tilde{G}_0(q) \left\{ e^{\ii q \hat{x}/\hbar} \overline{\rho}_{\FB} e^{-\ii q \hat{x}/\hbar} - \overline{\rho}_{\FB} \right\} \label{Eq:Flatband_dephasing} \\
&-\frac{2}{\hbar^2} \int_{-\infty}^{\infty} \!\! \dd q \, \tilde{G}_1(q) \int_{0}^{t} \!\! \dd t' \Big\{ e^{\ii v t' q/\hbar} e^{-\ii v t' (\hat{p}-p_1)/\hbar} \overline{\rho}_{\FB} \nonumber \label{Eq:Flatband_decay} \\
&\phantom{aaa} - e^{\ii q \hat{x}/\hbar} \overline{\rho}_{\DB} e^{-\ii v t' (\hat{p}-p_1)/\hbar} e^{-\ii q \hat{x}/\hbar} + h.c. \Big\} .
\end{align}
\end{subequations}
For the dispersive band component $\overline{\rho}_{\DB}$, one derives a similar evolution equation, with intraband dynamics as in Ref.~\cite{Gneiting2017disorder}. We remark that Eq.~(\ref{Eq:Flatband_evolution}) is not equivalent to Fermi's golden rule, which, while delivering asymptotic transition rates, remains ignorant about the intermediate dynamics.

Equation (\ref{Eq:Flatband_evolution}), which holds for arbitrary correlations and initial states, presents the basis for our analysis of the decay of the flatband states. It exhibits two components: A trace-preserving part (\ref{Eq:Flatband_dephasing}) describing the disorder-induced dephasing in the flatband channel, which causes a loss of coherence of the disorder-averaged state, along with a broadening momentum distribution. The second contribution (\ref{Eq:Flatband_decay}) captures the coupling into the dispersive band. As we show, the interplay between these two contributions ultimately limits the stability of flatband states.

\section{Decay into the dispersive band}

We first analyze the coupling of the flat into the dispersive band. We thus neglect for now the intrachannel dephasing (\ref{Eq:Flatband_dephasing}), corresponding to perfectly anticorrelated sublattice potentials, $\delta=-1$ in (\ref{Eq:Consistency_condition}) (we discuss the decay, however, for general $\delta$). Moreover, we assume that the dispersive-band state component is negligible, $\overline{\rho}_{\DB} \approx 0$. This is justified, because we consider the reliable storage sojourn of initial flatband states, i.e., before a significant dispersive-band component emerges. Also, any dispersive-band component propagates with velocity $v$, i.e., feedback into the flatband generally occurs remote from the initial flatband state location.

Equation~(\ref{Eq:Flatband_evolution}) can then be rewritten in momentum representation ($\overline{\rho}_{\FB}(p) = \bra{p}\overline{\rho}_{\FB}\ket{p}$), $\partial_t \overline{\rho}_{\FB}(p) = -\Gamma_t(p-p_1) \, \overline{\rho}_{\FB}(p)$, with the momentum-dependent decay rate
\begin{align} \label{Eq:exact_decay_rate}
\Gamma_t(p) = \frac{4}{\hbar^2} \int_{-\infty}^{\infty} \!\! \dd q \, \tilde{G}_1(q) \, t \, \sinc \left[ \frac{v t (q-p)}{\hbar} \right] .
\end{align}
The solution reads $\overline{\rho}_{\FB}(p) = \rho_{\FB, 0}(p) \, e^{-\overline{\Gamma}_t(p-p_1)}$, with $\overline{\Gamma}_t(p) = \int_{0}^{t} \dd t' \Gamma_{t'}(p) = \frac{2}{\hbar^2} \int_{-\infty}^{\infty} \!\! \dd q \, \tilde{G}_1(q) \, t^2 \, \sinc^2 \left[ \frac{v t (q-p)}{2 \hbar} \right]$. Assuming a finite correlation length $\ell$ further simplifies the decay in the limit $|v| t \gg \ell$:
\begin{align} \label{Eq:Approximated_decay_rate}
\overline{\Gamma}_t(p) = \frac{\pi t}{\hbar |v|} (1-\delta) G_0(p) .
\end{align}

We thus find a momentum-dependent decay of flatband states into the dispersive band, determined by the state's relative location w.r.t. the intersection, the transport velocity $v$ at the intersection, and the disorder characteristics.
As previously anticipated, this decay is absent if $\delta=1$, i.e., if the disorder potentials on the two sublattices are identical.

In the (unrealistic) limit of vanishing correlations, $C_{\aB \aB}(x) = C_{\bB \bB}(x) = C_0 \, \delta(x)$, we obtain $G_0(p) = \frac{C_0}{2 \pi \hbar}$, i.e., the decay happens homogeneously, irrespectively of the flatband state's position w.r.t.~the intersection. With Gaussian correlations, $C_{\aB \aB}(x) = C_0 \, e^{-(x/\ell)^2}$, we obtain $G_0(p) = \frac{C_0 \ell}{2 \sqrt{\pi} \hbar} \exp \left[ -\frac{1}{4} \left( \frac{p \ell}{\hbar} \right)^2 \right]$, i.e., (\ref{Eq:Approximated_decay_rate}) implies an exponential suppression of the decay for momenta $(p-p_1) \gg \hbar/\ell$. We remark that, in the short period before $t \approx \ell/|v|$, the exact rate (\ref{Eq:exact_decay_rate}) describes a transitional stage with decay spanning over a wider range of momenta; the impact of this stage is, however, generically small.

If the flatband state is (partly or fully) on resonance with the intersection, it rapidly begins to decay and spread in the dispersive channel, which, through backcoupling, results in spatial diffusion in the flatband channel. The momentum-dependent decay for finite $\ell$, on the other hand, suggests to store (sufficiently momentum-localized) states remotely (in momentum) from the intersection, in order to suppress their decay into the dispersive channel. However, as we show next, disorder-induced dephasing limits the temporal success of this strategy.

\section{Dephasing-mediated decay}

To assess the disorder-induced dephasing, we now neglect the dispersive-band coupling (\ref{Eq:Flatband_decay}), describing perfectly correlated sublattice potentials, $\delta=1$ (again, we keep $\delta$ general in the discussion). The remaining equation (\ref{Eq:Flatband_evolution}) is solved in position representation: $\bra{x} \overline{\rho}_{\FB}(t) \ket{x'} = \bra{x} \rho_{\FB, 0} \ket{x'} e^{-\overline{F}_t(x-x')}$, with
\begin{align}
\overline{F}_t(x) = \frac{t^2}{\hbar^2} \int_{-\infty}^{\infty} \!\! \dd q \, \tilde{G}_0(q) \left\{ 1 - \cos \left[ \frac{q x}{\hbar} \right] \right\} .
\end{align}
For Gaussian correlations, and with (\ref{Eq:Consistency_condition}), we then obtain $\overline{F}_t(x) = \frac{t^2 (1+\delta) C_0}{2 \hbar^2} (1-\exp[-(x/\ell)^2])$, i.e., the off-diagonal elements decay exponentially, causing a purity loss, while the diagonal elements remain unaffected.

More importantly, however, the dephasing comes with a distortion and broadening of the momentum distribution, as seen from the momentum variance, $\langle (\Delta \hat{p})^2 \rangle(t) = \langle (\Delta \hat{p})^2 \rangle_0 + \frac{t^2}{\hbar^2} \int_{-\infty}^{\infty} \!\! \dd q \, q^2 \tilde{G}_0(q)$, which, for Gaussian correlations, reads $\langle (\Delta \hat{p})^2 \rangle(t) = \langle (\Delta \hat{p})^2 \rangle_0 + \frac{(1+\delta) C_0}{\ell^2} t^2$, i.e., within our approximation, the momentum width increases linearly in time; the distribution in position space, however, remains unaffected.

In the general case, $-1<\delta<1$, we must consider both decay and dephasing. Moreover, we now include the contribution of the second intersection at $p_2=-p_1$. Recast in terms of the momentum distribution, Eq.~(\ref{Eq:Flatband_evolution}) then reads
\begin{align} \label{Eq:evolution_two_intersections}
\partial_t \overline{\rho}_{\FB}(p) =& -\sum_{j=1,2} \Gamma_t^{(j)}(p-p_j) \, \overline{\rho}_{\FB}(p) \\
&+ \frac{2 t}{\hbar^2} \int_{-\infty}^{\infty} \!\! \dd q \, \tilde{G}_0(q) \left\{ \overline{\rho}_{\FB}(p-q) - \overline{\rho}_{\FB}(p) \right\} , \nonumber
\end{align}
where $\Gamma_t^{(j)}(p)$ as in (\ref{Eq:exact_decay_rate}) (with $v_2=-v_1$). Note that the dephasing contribution manifests nonlocally here. Moreover, we remark that (\ref{Eq:evolution_two_intersections}) applies, similarly to (\ref{Eq:single_intersection_model}), also to other 1D flatband-intersection scenarios, possibly generalized to more than two intersections.

Based on our previous discussion, we should expect that, even if the initial state ($p_0=\langle \hat{p} \rangle$) is safely (i.e., decay-protected) located at $(p_0-p_j) \gg \hbar/\ell$ with momentum width $\langle (\Delta \hat{p})^2 \rangle_0 \ll (p_0-p_j)^2$, due to disorder-induced momentum broadening, the wavepacket extension reaches the intersection region of enhanced decay into the dispersive band, terminating the time span $\tau$ of the decay-protected sojourn. From the variance growth we estimate
\begin{align} \label{Eq:Lifetime_estimate}
\tau \lessapprox \frac{(p_0-p_1) \ell}{\sqrt{C_0 (1+\delta)}} ,
\end{align}
with $p_1$ the nearest intersection. In this sense, the presence of disorder introduces a lifetime for the reliable state storage in the flatband.

\section{Numerical test}

\begin{figure}[htb]
	\begin{center}
		\includegraphics[width=1.0 \linewidth]{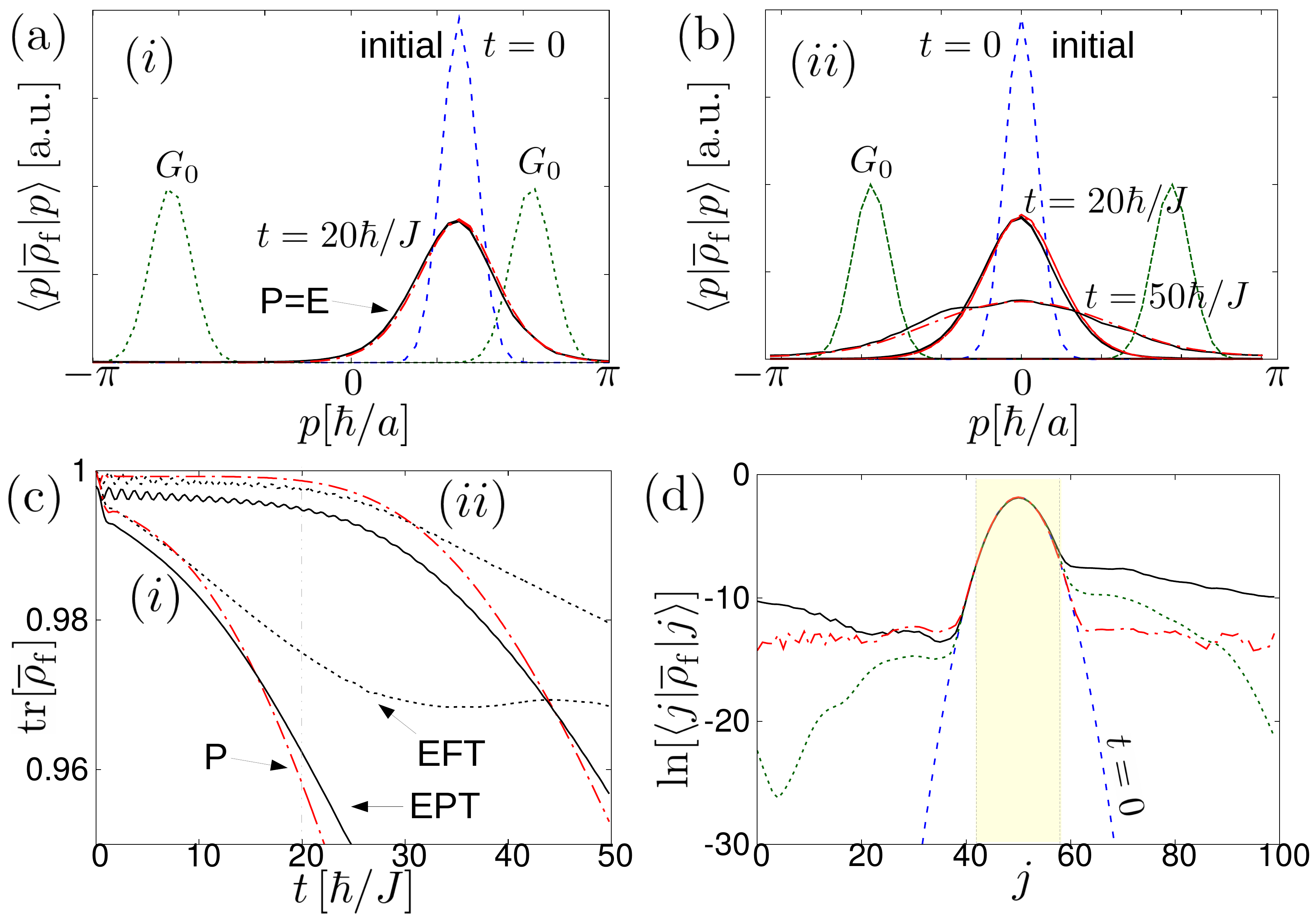}
	\end{center}
	\caption{\label{Fig:Flatband_decay} Disorder-induced decay of flatband states in the cross-stitch lattice. We compare the numerically exact evolution (black solid, E) with our prediction (\ref{Eq:evolution_two_intersections}) (red dash-dotted, P). Depending on whether the initial state (blue dashed) is (a)/(i) resonant with or (b)/(ii) detuned from the intersection [green dotted, for Gaussian correlations], the initial flatband state decays (c)/(i) steadily from the beginning, or (c)/(ii) not before the tails  reach the intersections [$t=20 \, \hbar/J$ in (b) and (c)]. While the overall decay into the dispersive band rapidly slows down [black dashed in (c), EFT=exact full trace], the flatband component exhibits ongoing diffusion [EPT=exact partial trace], resulting in the state's delocalization in (d). Depending on the state's relative position w.r.t.~the intersections and their transport velocities, this diffusion is symmetric [red dashed, (ii) at $t=20 \, \hbar/J$], or directional [green dotted, (i) at $t=10 \hbar/J$, and black solid, (i) at $t=20 \, \hbar/J$].}
\end{figure}

Figure~\ref{Fig:Flatband_decay} displays the time evolution for: the initial flatband state (i) partially overlapping with one intersection or (ii) residing in between the two intersections. In both cases we compare the numerically exact evolution in the cross-stitch model ($N=100$ unit cells, periodic boundary conditions, averaged over $K=200$ realizations) with our analytical prediction (\ref{Eq:evolution_two_intersections}). We use Gaussian correlations [the integral in (\ref{Eq:exact_decay_rate}) can then be solved analytically] with $W=0.5 J$ ($C_0=W^2/12$), $\ell=6 \lc$, and $\delta=0$. The intracell hoppings are (i) $\hp=1.0$ and (ii) $\hp=0.6$, along with the intersections (i) $p_{1,2}=\pm2.09 \, \hbar/\lc$ and (ii) $p_{1,2}=\pm1.88 \, \hbar/\lc$, and the velocities (i) $v=\pm3.46 \, \lc J/\hbar$ and (ii) $v=\pm3.82 \, \lc J/\hbar$. The initial flatband state is Gaussian, $\psi_0(x) \propto e^{-(x-x_0)^2/(2 \sigma_x^2)+i p_0 x}$, centered around $x_0=50 \lc$ with $\sigma_x^2=12 \lc^2$, and (i) $p_0=1.26 \, \hbar/a$ and (ii) $p_0=0$.

We find good agreement between our theory, within its range of validity, and the numerically exact results. In case (i) there is, due to the partial overlap of the initial state with the intersection, from the beginning a steady decay into the dispersive band. As anticipated, with a detuned initial state in (ii), the decay is delayed and sets in only after $\tau_{\rm exp} \approx 20 \, \hbar/J$. This delayed decay would be absent if $\delta=-1$, i.e., without intrachannel dephasing. In graphene, where $J \approx 2.8 \, {\rm eV}$ and 1D flatbands can, e.g., be found at the edges \cite{Yao2009edge}, the resulting lifetime estimate would, for above parameters, be $\tau \approx 5 \times 10^{-15} \, {\rm s}$. In 1D Lieb lattices of coupled micropillars \cite{Baboux2016bosonic}, with $J \approx 0.1 \, {\rm meV}$, one would obtain $\tau \approx 10^{-10} \, {\rm s}$. Note that (\ref{Eq:Lifetime_estimate}) overestimates the lifetime, $\tau \approx 80 \, \hbar/J$, as it is based on the variance, while the decay is sensitive to the tails reaching the intersections.

We remark that, in the numerical experiments, we measure the decay of the flatband state by taking the partial trace over the approximate carrier $[x_0-8 \lc, x_0+8 \lc]$ of the initial state in position space [yellow area in Fig.~\ref{Fig:Flatband_decay}(d)]. This is because, due to backcoupling (not modeled by our theory), the dispersive-band state partly reenters the flatband, however, due to propagation, remotely, this way contributing to the diffusive delocalization of the flatband state. The difference between partial and full trace then measures the fraction fed back into the flatband outside the carrier. We find a rapid slowdown of the overall decay into the dispersive channel, along with an ongoing diffusion of the flatband component, as predicted. The remaining deviations between our analytical predictions and our numerical results are explained by immediate partial feedback into the flatband before leaving the carrier, by reentering the carrier due to our finite, periodic lattice, by higher orders in $\hat{V}_{\varepsilon}$, and by discreteness effects. Note that the partial trace sets in slightly below $1$, due to the neglected initial-state fraction outside the carrier.

\section{Conclusions}

We specified a generic, disorder-induced decay mechanism in the interplay between flat and intersecting dispersive bands. We find that detuning flatband states from intersections delays their decay, limited by dephasing-mediated momentum diffusion. Backcoupling from the dispersive into the flat band eventually causes (potentially directional, i.e., ``chiral'') spatial diffusion of the flatband component. In this sense, disorder, while ineffective in isolated flatbands, when mediated by dispersive channels, delocalizes flatband states.

Whereas we exemplify our findings with the cross-stitch model, our theory holds for a wide range of 1D flatband scenarios with dispersive-band intersections, platform-independently. If intersections are absent (in the cross-stitch model, if $|\hp| \ge 2$), however, the identified mechanism is expected to be suppressed. Beyond their fundamental interest, we expect that our results are relevant, for instance, for the prospect of utilizing flatbands for state/information storage.

We stress that, in experimental implementations, depending on the platform, other factors, e.g., environmentally-induced decoherence and/or many-particle effects, can affect the evolution of the flatband states, possibly further reducing their stability. In this sense, the identified mechanism may serve as a baseline estimate on the stability of flatband states. Near-future experimental confirmations in existing platforms are conceivable \cite{Drost2017topological, Slot2017experimental, Li2017realization, Taie2015coherent, Ozawa2017interaction, Taie2017spatial, Nakata2012observation, Mukherjee2015observation, Vicencio2015observation, Kajiwara2016observation, Guzman2016experimental, Xia2016demonstration, Zong2016integrated, Baboux2016bosonic, Klembt2017polariton}. Generalizing the theory to 2D/3D, and including the evolution of dispersive-band components, could further illuminate the interplay of flat and dispersive bands. We expect that, in 2D or 3D, a similar mechanism prevails, i.e., the detuning from the closest dispersive-band intersection delays the onset of the diffusion process.

\section*{Acknowledgments}

We thank Alexander Rozhkov and Daniel Leykam for helpful discussions. F.N. is partially supported by the MURI Center for Dynamic Magneto-Optics via the Air Force Office of Scientific Research (AFOSR) (FA9550-14-1-0040), the Army Research Office (ARO) (Grant No.~W911NF-18-1-0358), the Asian Office of Aerospace Research and Development (AOARD) (Grant~No.~FA2386-18-1-4045), the Japan Science and Technology Agency (JST) (the ImPACT program and CREST Grant No.~JPMJCR1676), the RIKEN-AIST Challenge Research Fund, the Japan Society for the Promotion of Science (JSPS) (JSPS-RFBR Grant No.~17-52-50023 and JSPS-FWO Grant No.~VS.059.18N), and the John Templeton Foundation. Z.L. acknowledges support from a JSPS postdoctoral fellowship.

\bibliography{literature}

\end{document}